**The case for biophysics super-groups in physics departments**

Bart W. Hoogenboom[1,2] and Mark C. Leake[3]

[1] Biological Physics Research Group, Department of Physics & Astronomy, University College London, Gower Street, London, WC1E 6BT, United Kingdom.

[2] London Centre for Nanotechnology, University College London, 17-19 Gordon Street, London, WC1H 0AH, United Kingdom

[3] Biological Physical Sciences Institute (BPSI), Departments of Physics and Biology, University of York, York, YO10 5DD, United Kingdom.

For correspondence email: b.hoogenboom@ucl.ac.uk or mark.leake@york.ac.uk

*Increasing numbers of physicists engage in research activities that address biological questions from physics perspectives or strive to develop physics insights from active biological processes. The on-going development and success of such activities morph our ways of thinking about what it is to 'do biophysics' and add to our understanding of the physics of life. Many scientists in this research and teaching landscape are homed in physics departments. A challenge for a hosting department is how to group, name and structure such biophysicists to best add value to their emerging research and teaching but also to the portfolio of the whole department. Here we discuss these issues and speculate on strategies.*

**What is bio(logical) physics?**
But a few centuries ago, physics and biology were primordial components of a single discipline initially called natural philosophy and later natural science, prior to bifurcating along different intellectual paths. By the early 20$^{th}$ century, however, aspects of physics and biology were reunited, exemplified in D'Arcy Wentworth Thompson's mechanical description of biological growth and shape[1]. In the 1950s, biophysics research pioneered major developments in physiology and structural biology: exemplified by the Hodgkin-Huxley model that describes the propagation of electrical signals in neurons[2], and by the discovery of the DNA double helix by Watson, Crick, Franklin, Wilkins and others, based on X-ray diffraction experiments[3–5].

Since the 1950s, experimental and theoretical techniques from physics rapidly developed to address a range of biological questions across extensive length and time scales: research on populations of organisms in macroscale ecosystems, as well as – at the nanometre length scale – research on individual biomolecules[6]; biophysical phenomena at femtoseconds time scale through to biological processes evolving over many years. Besides new insights into biology, these developments led to new physics not necessarily coupled to questions relevant for living objects, such as a 'moving version of the Heisenberg model' in the of context active biological matter (exemplified by analogies between quantum coupling in magnetic materials and the spatial patterns of flocking behaviour of populations of flying birds[7]).

In our view, biological physics – which we denote simply as biophysics – encompasses all these research types, be they inspired or motivated by biological questions, where the physics component can lie in the nature of the (experimental/theoretical/computational) tools that are used and/or in the type of science that is generated.

**Is biophysics physics?**
Interestingly, when physical approaches are really successful in biology, they are often absorbed by other disciplines: The above-mentioned Hodgkin-Huxley model and DNA double helix structure, both strongly grounded in physics, were awarded Nobel Prizes in Physiology or Medicine (1963 and 1962, respectively). More recently, in spite being rooted in physics, developments of super-resolution fluorescence microscopy and cryo-electron microscopy were awarded Nobel Prizes in Chemistry (2014 and 2017). Hence, in response to the common misconception that biophysics is simply 'not physics', one may – hyperbolically – retort that the 'less physics' biophysics appears to become, the more important it is.

That said, it is not difficult to find examples of outstanding biophysics that is firmly and unambiguously categorised as physics. One example is the pioneering work of Pierre Gilles de Gennes, awarded the Nobel Prize in Physics in 1991 for studying order phenomena in simple systems in a way that could be generalized to more complex forms of matter, after extending Sam Edwards's seminal work. He developed new polymer physics theories, which involved reptation and branching, steered in no small part by observations of biological polymers, resulting in invaluable biological insights. There is Steve Chu, Nobel Prize winner in Physics in 1997 for his work on cold-atom trapping, who later applied laser trapping technologies towards understanding biomolecules, resulting in important biological insights into the nature of mechanical relaxation of DNA molecules. And more recently Steve Block has used innovative single-molecule biophysics techniques to map out the free energy landscape for nucleic acids – certain forms of these molecules exhibit a wide range of conformational microstates. This work is a single-molecule experimental application of the Jarzynski equality, one of the most important theories of modern statistical mechanics. The physics involved in all three examples is fundamental, but the results have been enormously influential towards understanding biology.

More playfully, we note that the diffusion equation, an immensely important equation in biophysics, is equivalent to a Schrödinger equation in imaginary time, and there is little doubt that quantum-mechanical research involving the Schrödinger equation is physics. In short, biophysics is an important part of physics, as has been firmly and repeatedly articulated and illustrated in past and present[8–11].

**Biophysics in physics departments**
While biophysics research can typically be found across university departments and faculties (e.g., in chemistry, biology, physiology), there are numerous reasons, both scientific and practical, for having a strong biophysics component in a physics department in particular, as outlined below.

Firstly, the modern research landscape is highly interdisciplinary in nature, much of it operating at the interface between the physical and life sciences; and so are the demands of many emerging high-tech industries (*i.e.*, future employers for physics students). Physics departments are now responding to these new demands by incorporating biophysics activities in research and teaching.

Secondly, funding bodies increasingly recognise the need to support biophysics research, often via targeted calls involving joint investment from funding bodies with portfolios in engineering/physical sciences and biological/biomedical sciences: By having depth and breadth of biophysical expertise, physics departments are in a better position to develop competitive proposals.

Thirdly, in the context of teaching physics at university, there are great benefits from being able to pool into biophysics expertise[12]. Undergraduate physics concepts can be vividly illustrated by examples from the life sciences: The overdamped harmonic oscillator model can be applied to muscle contraction or tetanus; the diffraction of waves underpins the limits of spatial resolution with which we can investigate the living cell; knowledge of electrical circuits is needed to understand the propagation of signals along nerve cells; even quantum physics has its uses in biology, e.g., to describe photosynthesis; and one can introduce many concepts of statistical physics with biological applications.

**Biological physics versus condensed matter physics**
A common route to establishing biophysics in a physics department is to coral biologically relevant activity into a condensed matter physics super-group of some form (we here define a "super-group" as a gathering of different principal investigators and their labs/groups). After all, living matter is a form of condensed matter. This has often been seen as the best fit but can create challenges at several levels in the instance of one or only a few investigators being engaged in biophysics. As biophysics grows, departmental discussions in some cases involve sentiments such as 'biophysics cannot be a super-group because there are too few faculty members in the department', 'we cannot break up the current structure because it will disrupt the recycling of departmental funds to faculty members', 'it's not the right time to change the shape of the department', or 'there is not sufficient new physics to justify a biophysics super-group on equal basis as others'. Such a debate is often followed by a compromise in the form of new sub-groupings of biophysics: 'soft/active matter', 'biomaterials', 'biological physics/physical biology', or sematic variants thereof. Alternatively, one may change the name of a condensed matter super-group to suggest a greater complexity.

There is a general risk of pooling various emerging physics disciplines as generic 'condensed matter physics' as soon as they involve aspects of matter in a condensed phase. The identity of a large condensed matter physics super-group can become confused, because of difficulties in articulating clear overarching themes that cover, e.g., quantum computing, topological insulators and biophysics. This problem is rather common for – but of course not unique to – modern condensed matter super-groups. In such cases, the 'condensed matter' label is simply inadequate to describe the range of intellectual diversity. It may satisfy internal administrative needs but can misrepresent the group to the outside world.

Such a misrepresentation does not support the emerging identity of biophysics in physics departments, and potentially stifles growth. On the other hand, the phrase 'biophysics' is loaded with pre-conceptions to its applications in the life sciences, among others by its often being paired up with structural biology. However, it is possible to reclaim the word – or its extended version, 'biological physics' – to represent a broad, interdisciplinary community populated by researchers from both physical and life sciences backgrounds but converging on similar scientific aims.

**Inventory of biophysics (super-)groups in UK physics departments**
Given the various routes to support, channel and represent biophysics, we investigated how collective biophysics activities are organised in UK physics departments. By data-mining of all listed UK physics department websites and by collecting straw-poll responses from senior biophysics researchers hosted by these departments, we have categorised departmental biophysics groupings as follows (Figure 1):
- Super-group: Interdisciplinary physical/life sciences is core to a collection of more than one individual research team; this collection of teams has a recognised autonomy for managing small to medium budgets within the department to the same extent as other recognised major groupings.
- Virtual group: This has the outward appearance of super-group but in reality is managed by one or more other super-groups (often 'condensed matter physics' or equivalent) for budgetary/administrative matters.
- No collective grouping: There is no cohesive super-group, because there is only one biophysics team in the department or, if more than one, then these teams do not perceive themselves as a collective structure.
- No biophysics: There is no research team in the department whose research/teaching portfolio comprises at least 50% biophysics.

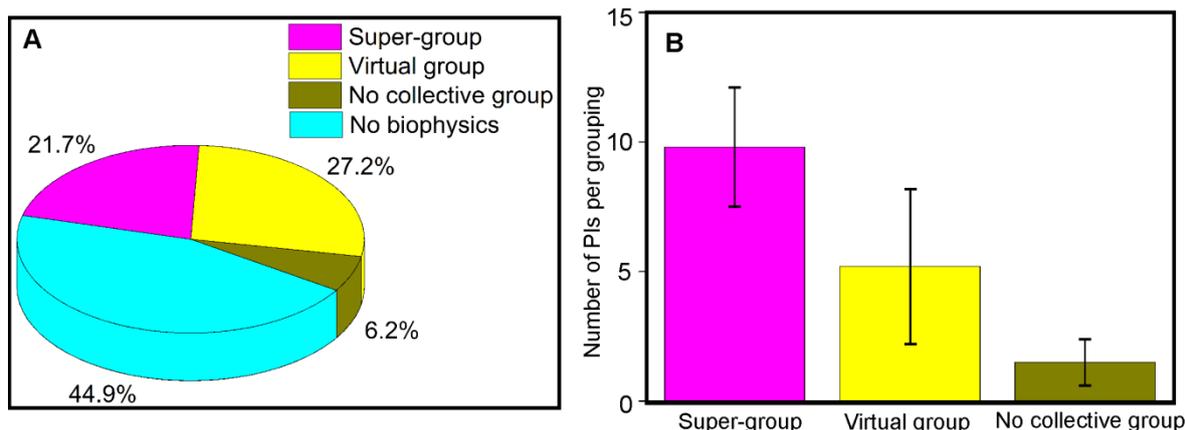

**Figure 1.** Biophysics groupings in UK university physics departments. (A) Proportion of UK physics departments that either have an autonomous biophysics super-group, virtual group, no collective group, or no biophysics at all (see main text for further description). (B) Histogram showing mean number of principal investigators per biophysics grouping for the three categories of (A) that comprise non-zero biophysics components (error bars refer to standard deviations). Data acquired from accessible websites from 49 listed 'Physics and Astronomy' departments in the Complete University Guide 2018 [13], checked against straw-poll responses from 19 senior UK biophysics researchers.

An important result (Fig. 1A) is that over half (27 from 49) of UK physics departments have some biophysics presence, estimated from website data to comprise 800-900 active researchers (PhD students, postdocs/fellows and faculty) at the time of writing. That said, a significant minority of UK physics departments (45%) have no significant biophysics presence. Of the biophysics groupings, the largest category, roughly half of all departments with biophysics presence, is that of a virtual group; 39% of these departments have biophysics super-groups; and 11% of these had no collective grouping. Predictably, there was a trend in the number of principal investigators (PIs) in each category (Fig. 1B), with a mean $10 \pm 2$ PIs per grouping (standard deviation, number of physics departments $n = 6$) for super-groups, $5 \pm 3$ ($n = 10$) for virtual groups, and $1.5 \pm 0.9$ ($n = 11$) where no collective group was present.

These data suggest that more than 60% of biophysics academics in UK physics departments are not currently associated with an autonomous biophysics super-group. Also: (i) in four physics departments there exist two separate virtual groups in the remit of biophysics, and (ii) of the 11 physics departments without collective biophysics groups, three have more than one research team (*i.e.*, there are isolated biophysics teams not structured into a collective group). There are also pockets of biophysicists with a physics background in life sciences departments, in engineering and chemistry departments and in a number of virtual interdisciplinary centres, as well as in interdisciplinary research centres funded by biology and/or biomedicine funding bodies, e.g. Medical Research Council (MRC) funded laboratories themed in molecular/cell biology and general medical sciences at Cambridge, University College London and Imperial College London, the Biotechnology and Biological Sciences Research Council (BBSRC) funded John Innes Centre, and various Wellcome Trust (WT) funded research centres, not to mention the Francis Crick Institute involving University College London, Imperial College London and King's College London which is co-funded from the MRC, WT and Cancer Research UK (CRUK). An important role of biophysics super-groups in physics departments is to reach out to these other pockets of biophysics research activity.

Other qualitative responses emerged from the straw-poll, reflecting some uncertainty about 'what biophysics is' at senior management levels of UK physics departments. Two example quotes from senior biophysics researchers are: *'Until recently biological physics was barely recognised at all… fair to say that the department doesn't really know how to handle biological physics'* and *'In a sense we aren't managed at all, just left alone'*.

**Case studies of biophysics groupings in physics departments**
Two case studies, from the physics departments of the University of York and University College London, illustrate how biophysics is positioned in different ways within different universities.

At the University of York, biophysics activities are gathered in a virtual group. In its Department of Physics, biophysics activities increased significantly in 2013 with the recruitment of a new chair, and subsequent recruitment of a lecturer and several early career staff, with a total number of 10 current independent research fellows and academics whose core activities involve biophysics. Most biophysics is pooled as part of a large condensed matter physics super-group comprising 25 academics. This super-group covers five overlapping themes of nanomaterials, photonics, quantum science, spintronics & magnetism, and biophysics & biomaterials. There are also links to biophysics activities in other departments through a cross-disciplinary network of researchers called the Biological Physical Sciences Institute, funded by Departments of Physics, Biology and Chemistry. An autonomous biophysics seminar series in the Department of Physics has increased in popularity over the past few years, also beyond the condensed matter physics super-group, to capture interest from other existing super-groups.

In the Department of Physics & Astronomy at University College London, biophysics activities were initially (from 2009) gathered in a virtual grouping of faculty from its Atomic, Molecular, Optical and Positron Physics and from its Condensed Matter and Materials Physics super-groups. In 2014, this virtual group was transformed into a super-group in Biological Physics, still smaller than but administratively on par with the other four research super-groups in the department. At present it includes 13 tenured academics. Of these, three are not employed by the department, but affiliated for other reasons. Some members have retained a partial affiliation with another research super-group, though the intention is to gradually phase out such joined affiliations. The Biological Physical super-group has its own budget, which is allocated from the departmental budget based on its total number of full-time academic staff. The Biological Physics group is also a key player in the university's Institute for the Physics of Living Systems, a virtual centre which gathers a large biological physics community across departments and faculties.

**Arguments for biophysics super-groups**
Provided that the number of staff is sufficient to justify the formation of an administrative entity such as a biophysics super-group, this creates a formal path for input in departmental strategy, to ensure that biophysics activities are properly taken into account and where appropriate strengthened. It also provides a formal framework for mentoring, for mutual support, and for cohort formation of graduate students, with the advantage that this is provided by colleagues/students working in a related research field.

Super-group formation enhances the visibility of the biophysics research activities of a department, for students, for potential (biophysics) recruits, for potential academic and industrial partners, and for funders. Increased visibility is also important because the recognition of biophysics as a field by undergraduates lags behind in the UK compared with other countries such as France and Germany. This representation function can in part be achieved by virtual groupings, although this is at the risk of dilution in the presence of multiple network structures that can be present at a university.

There are also pragmatic financial reasons to consider models that enable biophysics to grow into research super-groups. Business plans vary across different departments but generally involve recycling of overhead income from external grants back to group leaders, typically small sums of a few £k per year. However, within a biophysics super-group, these funds can be routed into nurturing biophysics activities directly, for example networking, seminar series, funds for project students, and travel to biophysics conferences. Although these are small funds in comparison to external grant income, they can sustain the general biophysics concept inside a physics department. In some cases, overheads recycling extends to higher amounts, and pooling these enables dedicated technical/administrative staff to be hired, with more tangible benefits to sustaining biophysics.

Networking funds are particularly essential in interdisciplinary research, as its success strongly depends on encounters between researchers typically based in different departments.

When under the umbrella of a non-biophysics super-group, biophysicists run a risk of losing out in the overheads balancing act. This structure is likely to prove increasingly unpopular as greater investment is made into biophysics research: The Engineering and Physical Sciences Research Council (EPSRC) describes biophysics as one of its growth areas, and there are new cross-council initiatives that increasingly support biophysics activities, such as the Global Research Fund, Antimicrobial Resistance, Multidisciplinary Project Awards from CRUK, and Technology Touching Life. By its multidisciplinary nature, this income can be tapped from multiple funding bodies: This may prove pragmatic in the event of departmental financial stress tests, a bet-hedging strategy more prudent than putting all of one's eggs into one funding body basket. By taking advantage of more collective outputs, a diverse biophysics super-group may enhance its chances of winning major interdisciplinary grants across a wide range of funding sources compared with less collaborative research consortia.

Recent independent reports highlight the increase of interdisciplinary in the UK. The British Academy appraised the cultural challenges within UK academia[14]; UK research councils were reviewed by Sir Paul Nurse[15], and the research excellence framework (REF) was discussed by Lord Nicolas Stern[16] as evidence of how interdisciplinary science taps into key remits of several research funding councils exceptionally well, but is hampered by organisational and administrative structures of the councils and academic institutions. At the level of physics departments, biophysics super-groups improve the level of interdisciplinary cohesion: They work towards aligning with the recommendations in these reports for developing structures that are more robust with regards to nurturing interdisciplinarity.

**Conclusion**
There are several important reasons for developing strong biophysics in physics departments. However, over half of UK physics departments either still do not have any biophysics activity or have a biophysics presence that is hidden behind historic structures of research and teaching. Based on our analysis of the organisation of biophysics in UK physics departments, we conclude that there is scope for immediate improvement as follows:
- The four physics departments that have more than one virtual biophysics group could benefit from consolidating their biophysics activities into a super-group to improve visibility and cohesion.
- The three physics departments that have several PIs who are not part of a collective group structure might similarly benefit from consolidating into at least a virtual group.
- The intersection between super- and virtual groups in terms of numbers of PIs lies at 7-8 per group. In other words, virtual groups with at least 7 separate research teams might qualify as having 'critical mass' for a super-group, relevant currently to three virtual groups in the UK.
- Taken together, it would be feasible for 12 UK biophysics super-groups to exist given the restructuring suggested above, double the number a present, a far more visible identity and force for change.

In spite of an active research community, the UK does not have the international visibility as a hub for biophysical research it deserves, mostly because of a lack of structure. One way to improve the national visibility of biophysics, in addition to fostering the growth of more biophysics super-groups, is for biophysicists across the biology-physics interface to become more unified. In the UK, this is exemplified by the longevity of regular international meetings and focused workshops such as those organized by the Biological Physics Group (BPG) of the Institute of Physics, including Physics Meets Biology; the Physics of Living Matter Symposium organised by the University of Cambridge and University College London; and several more events organized by the Physics of Life network[17] and the British Biophysical Society (BBS)[18]; and recently (2017) by the success of the Joint 19th International Union of Pure and Applied Biophysics (IUPAB) and 11th European Biophysical Societies' Association (EBSA) Congress in Edinburgh. This event drew thousands of the world's best biophysicists to the UK thanks to combined efforts of the BBS and the BPG. The BBS and BPG have traditionally represented the UK biophysical interests from more polar perspectives of biology and physics, respectively. However, this successful convergence in Edinburgh illustrated a unified feature of biophysics, which can equally capture biology and physics. Unity at a level of two national societies may offer a valuable template for physics and biology departments to follow; namely, that a biophysics super-group can, and perhaps should, capture expertise from physics and biology departments, for example though establishing joint academic cross-departmental appointments.

Ultimately, it is in the crowd, with shared identity and purpose, that things can change. A collective moment can result in real change, but it is important that a crowd does not become a mob; it needs structure, and accepted routes of engagement. It is very difficult to change things for the better as a single individual: Departmental super-groups offer a potential way forward to build a strong national biophysics community for the future in the UK.

**Acknowledgements**
The authors thank Ewa Paluch (University College London) for providing helpful comments and suggestions. ML acknowledges funding from BBSRC grant BB/R001235/1 and MRC grant MR/K01580X/1.